\documentclass[reprint,superscriptaddress,
 amsmath,amssymb,
 aps,
]{revtex4-2}
\usepackage{graphicx}
\usepackage{dcolumn}
\usepackage{bm}
\usepackage{tikz}
\usetikzlibrary{arrows.meta, positioning}
\usepackage{xr}
\usepackage{xcolor}
\usepackage[normalem]{ulem}
\definecolor{MC}{RGB}{0, 105, 60} 
\usepackage{bibunits}
\begin{document}

\title{Delayed Formation of Landau Polaritons in Phase-Resolved THz Spectroscopy}

\author{Noureddine Charrouj}
\affiliation{Centera Labs, Institute of High Pressure Physics, Polish Academy of Sciences, Warszawa, Poland}

\author{Nischal Gajurel}
\affiliation{School of Physics, Engineering and Technology, University of York, United Kingdom}

\author{Yurii Ivonyak}
\affiliation{CENTERA Labs, Institute of High Pressure Physics, Polish Academy of Sciences, Warszawa, Poland}
\affiliation{CENTERA, Cezamat, Warsaw University of Technology, Warszawa, Poland}

\author{Dmitriy Yavorskiy}
\affiliation{Centera Labs, Institute of High Pressure Physics, Polish Academy of Sciences, Warszawa, Poland}
\affiliation{Centera, Cezamat, Warsaw University of Technology, Warszawa, Poland}

\author{Vladimir Y.\ Umansky}
\affiliation{Weitzman Institute of Science, Rehovot, Israel}

\author{Jerzy Łusakowski}
\affiliation{Faculty of Physics, University of Warsaw, Warszawa, Poland}

\author{Wojciech Knap}
\affiliation{Centera Labs, Institute of High Pressure Physics, Polish Academy of Sciences, Warszawa, Poland}
\affiliation{Centera, Cezamat, Warsaw University of Technology, Warszawa, Poland}

\author{Marcin Białek}
\affiliation{Centera Labs, Institute of High Pressure Physics, Polish Academy of Sciences, Warszawa, Poland}
\email{marcin.bialek@unipress.waw.pl}

\date{\today}

\begin{abstract}
Strong light–matter coupling gives rise to polaritons through coherent and periodic energy exchange between electromagnetic cavity fields and material excitations. While this interaction is typically inferred from spectral mode splitting, its dynamics remain largely unexplored. Here, using phase-resolved terahertz time-domain spectroscopy, we observe vacuum Rabi oscillations of Landau polaritons formed by coupling the cyclotron resonance in a GaAs/Al$_{0.36}$Ga$_{0.64}$As two-dimensional electron gas with Fabry–Perot cavity modes.
By employing cross-polarized spectroscopy and magnetic-field differential, we resolve the temporal beating of the cyclotron resonance oscillations.
Supported by a quantum electrodynamic model, we show that the vacuum Rabi oscillations start after a delay corresponding to one cavity round-trip time. This demonstrates that the strong-coupling regime sets up only after the formation of the cavity mode field. Our results provide direct insight into the dynamics of hybrid light–matter states in the THz regime.
\end{abstract}


\maketitle
\begin{bibunit}
Strong light–matter coupling is a coherent and periodic energy exchange between cavity electromagnetic fields and excitations of matter faster than the decay times of these modes \cite{Khitrova06}. This periodic energy exchange manifests as beating in matter and in optical modes, known as Rabi oscillations. Beating in the time domain is reflected in the frequency domain as the Rabi splitting into upper and lower polariton states.
Polaritons are attractive for many applications \cite{Basov25}, including radiation sources and detectors \cite{Kockum19, Solnyshkov21, Xiang24}, or to modify the ground state of matter \cite{Baydin25}. The coupling of a photonic field with $N$ resonators is enhanced by a factor of $\sqrt{N}$, known as the Dicke cooperativity \cite{Yahiaoui22}.
When the splitting is a substantial fraction of the oscillators' frequencies (typically 10\%), the coupling enters the ultrastrong regime \cite{Kockum19, Forn-Diaz19}. 

Despite this interest in polariton studies, the Rabi oscillations are usually observed indirectly in frequency spectra as the splitting into upper and lower polariton modes \cite{Weisbuch92, Yoshie04, Johansson06, Dominici14, Zhang14, Dunkelberger16, Silva20}.
This is because most spectroscopy techniques are phase-insensitive and measure only the envelope of the beating pattern.
There is a handful of phase-resolved reports on Rabi oscillations; for instance, in mid-IR intersubband polaritons \cite{Gunter09}, in a THz pump-optical probe system was used to demonstrate anomalous temporal growth of the magnetization oscillation amplitude in a bulk antiferromagnetic sample \cite{Blank23}, related to the weak magnon-photon coupling. For phonons, Rabi oscillations have been demonstrated using the THz time-domain spectroscopy (THz-TDS) method and a tunable cavity \cite{Jarc22}. This method offers phase-sensitive measurements of the electric field of radiation thanks to the time resolution of about 0.1 ps, much shorter than the typical wave period of the order of 1 ps (1 THz). 
Therefore, THz-TDS gives the possibility to study the dynamics and formation of phase-resolved beating of the modes caused by the Rabi oscillations of a period longer than a few ps. 
In this work, we study the time-domain formation of light-matter coupling in high-order Fabry-Perot modes, while typically studied THz cavities operate at the fundamental mode \cite{Gunter09, Halbhuber20, Morhinweg21}
Therefore, in this study, the light cycle is a few times shorter than the cavity cycle, allowing us to clearly observe time-resolved polariton formation. 

In the THz frequency range, strong coupling can be realized with various solid-state excitations, such as phonons \cite{Jarc22, Kim23, Bialek25}, magnons \cite{Huebl13, Zhang14, Tabuchi15, Bai15, Bialek21, Kritzell24}, intersubband transitions \cite{Dini03, Jeannin19}, and excitations of a two-dimensional electron gas (2DEG) \cite{Geiser12, Paravicini-Bagliani17}. Among these, the cyclotron resonance (CR)---the transition between quantized Landau levels in semiconductor electron bands---stands out due to its inherently strong interaction with electromagnetic radiation and its tunability by an external magnetic field \cite{Zhang14PRL, Muravev25}.
Different types of optical cavities were shown to exhibit strong coupling with the CR, including metasurfaces \cite{Scalari12, Scalari13, Rajabali21, Morhinweg21, Mornhinweg24}, Fabry-Perot cavities \cite{Muravev11, Zhang16, Li18, Shuvaev21, Tay25, Mavrona21, Yavorskiy26}, and single meta-atoms \cite{Jochl26}.
When the CR strongly interacts with the cavity field, the resulting states are called Landau polaritons, characterized by a broken time-reversal symmetry \cite{Andberg24}.

We study the light-matter coupling between the CR and high-order Fabry-Perot modes formed in the sample slab. The vacuum Rabi splitting scales as $p\sqrt{N f_m / V_m}$ \cite{Khitrova06}, where $ N$ is the number of oscillators, $p$ is the transition dipole moment, $f_m$ is the $m$-th cavity mode frequency, and $V_m$ is the cavity volume, which is proportional to the substrate thickness. The $m$-th Fabry-Perot mode at frequency $ f_m$ crosses with the CR at a magnetic field $B_m=f_mm^*/e$, where $m^*$ is the conduction band electron effective mass, and $e$ is the elementary charge. In the case of the CR, $p\propto 1/\sqrt{B}$; thus, the vacuum Rabi splitting is the same for all successive Fabry-Perot cavity modes.

Our samples are GaAs/Al$_{0.36}$Ga$_{0.64}$As triangular quantum wells hosting a 2DEG with an electron density $n_s\sim3\times10^{11}$ cm$^{-2}$ and an electron mobility $\mu>10^5$ cm$^2$/Vs at 4~K.
Since the substrate thickness controls cavity modes and the strength of the light–matter coupling, we study two samples of different thicknesses. The sample referred to as the "thick-substrate" has the original thickness of $383\pm2~\mu$m, while the "thin-substrate sample" is thinned down to $131\pm2~\mu$m by mechanical polishing.

Using THz-TDS, we measure reflection spectra at an incidence angle of $\sim8^\circ$ of samples at 10~K (see Supplementary Materials Fig.\ S1)
We independently control the polarizations of the incident and the reflected beam with two polarizers. 
We measure co-polarized reflection time-domain traces $R_x$ with the polarizers parallel. 
In cross-polarized measurements of $R_y$, the two polarizers are oriented at $90^\circ$ to each other \cite{Arikawa11, Zhang2014}.

\begin{figure*}
    \centering
    \includegraphics[width=\linewidth]{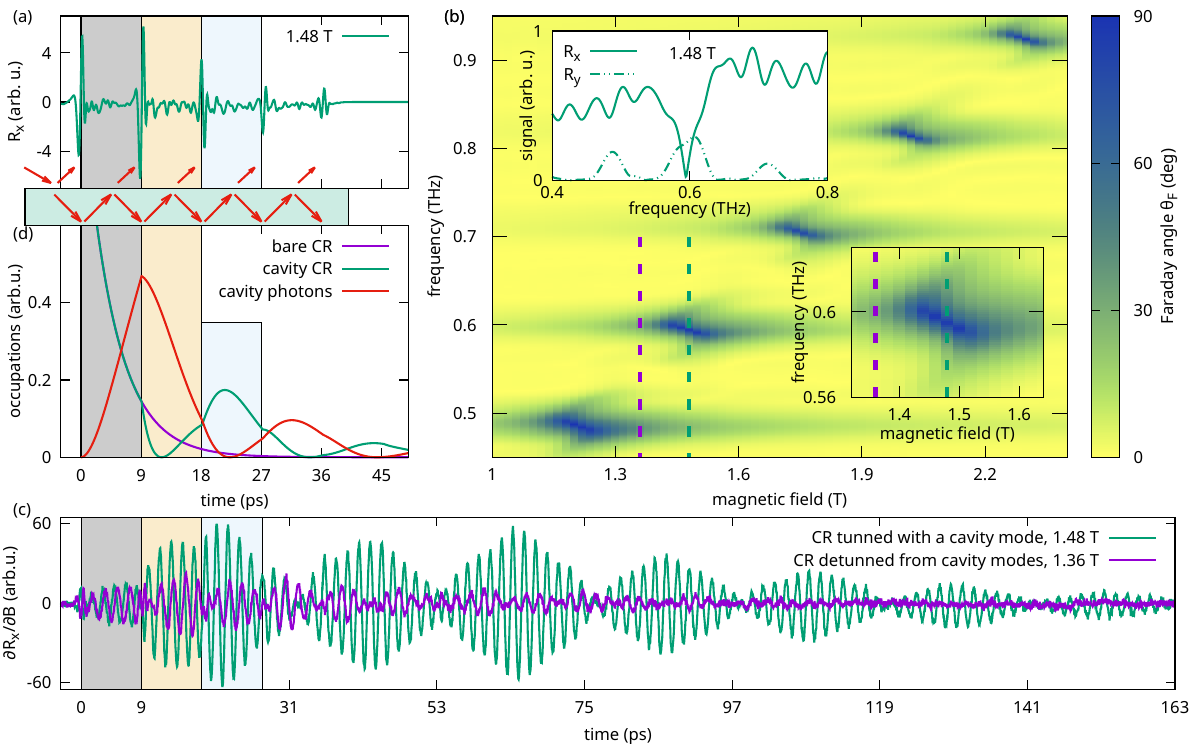}
    \caption{Sample with thick substrate.
    In the segments (a), (b), and (d), the time scale is the same; the gray, yellow, and blue areas represent time spans following three subsequent THz pulse echoes.
    (a) Time-domain trace of co-polarized reflection $R_x$ at $B=1.48$~T. In the schematic below, the red arrows symbolize the propagation of THz pulses in the slab (the green rectangle). Multiple internal reflections produce a train of pulses, as visible in the time-domain trace.
    (b) Experimental Faraday rotation angle $\theta_F$ as a function of magnetic field. Arrows and dashed lines mark the magnetic fields of the time-domain traces presented in other panels.
    The inset on the left shows frequency-domain spectra obtained from traces in (a) and (b). 
    The inset on the right shows an enlarged region of the CR interaction with the 6th cavity mode.
    (c) $B$-field-differential time-domain reflection traces of the thick-substrate sample. The ticks of the timescale mark the nodes of the vacuum Rabi oscillation.
    (d) Quantum-electrodynamic-model-based occupancies of the bare CR mode on top of semi-infinite GaAs substrate (violet), CR mode in the thick-substrate sample (green), and the cavity mode (red).
    }
    \label{fig:cross-thick}
\end{figure*}

In our samples, the reflections of the incident THz pulse in the sample substrate create a train of THz pulse echoes separated by the cavity period, i.e.\ twice the time-of-flight through the sample. In Fig.~\ref{fig:cross-thick}a, we demonstrate such a result showing a train of pulses with about 9 ps period, observed for the thick-substrate sample.
Such a sequence of pulses is the time-domain picture of the Fabry-Perot cavity spectrum. 
All of these THz pulse echoes arriving at the 2DEG excite the CR, which is expressed as a small oscillation following the pulses.

Before showing the vacuum Rabi oscillations in the time domain, we need to identify the conditions under which we observe the strong coupling. 
For that, we analyze frequency-domain reflection spectra in co- $|R_x(f)|$ and cross-polarization $|R_y(f)|$, obtained by the fast Fourier transform of the respective time-domain traces. We show an example of such spectra in the left inset of Fig.\ \ref{fig:cross-thick}c.
Subsequently, we calculate the Faraday rotation angle spectra as $\theta_F=\arctan(|R_y(f)|/|R_x(f)|)$. This procedure also naturally normalizes the spectra.
The evolution of the Faraday angle spectra with magnetic field (Fig.\ \ref{fig:cross-thick}b) reveals zigzag patterns formed due to the strong coupling of the CR with a few subsequent Fabry-Perot modes \cite{Li18}.
This pattern forms because the CR couples to both co- and counter-rotating components of cavity mode fields, while we probe the samples using linearly polarized THz pulses. The upper and lower branches of the zigzag arise as a result of the strong coupling of the CR with the co-rotating cavity field. The middle mode is related to the breaking of the rotating-wave approximation and the vacuum Bloch-Siegert shift, which arises from the weak coupling between the CR and the counter-rotating component of the cavity mode field.
Remarkably, the Faraday rotation angle reaches 90 degrees \cite{Suresh25}. This happens at frequencies for which $|R_x(f)|$ drops to very low values, while $|R_y(f)|$ reaches maximal values. Thus, under these conditions, coinciding with Landau-polariton modes, the reflected radiation is polarized in a direction perpendicular to the incident polarization.

Knowing the conditions at which the system is in the strong coupling regime, we show the Rabi oscillations by calculating the magnetic-field derivative of time-domain traces, i.e.~by subtracting traces measured at slightly different magnetic fields $\partial R_x/\partial B = R_x(t,B+dB)-R_x(t,B)$, $dB=0.02$~T. This method reveals the portion of the reflection signal that is dependent on the magnetic field. 
In Fig.\ \ref{fig:cross-thick}c, we show two $\partial R_x/\partial B$ traces; the green line shows the beating of the CR oscillations observed at $B=1.48$~T, i.e., vacuum Rabi oscillations caused by the strong light-matter coupling; while the violet line shows the result when the CR was detuned from cavity modes (for more spectra see Supplemental Material Fig.\ S4). In the tuned case, we observe clear vacuum Rabi oscillations with nodes at times marked with the time scale ticks. In contrast, the detuned case shows a significantly smaller overall amplitude of CR, which is modulated with a cavity period (9 ps) that is much shorter than the vacuum Rabi oscillations period (22 ps). Color rectangles in the background mark the cavity periods, and the time scale in Fig.\ \ref{fig:cross-thick}d is the same as in panels (a) and (d).

Crucially, in Fig.\ \ref{fig:cross-thick}c during the first cavity cycle at $t=0-9$ ps (gray rectangle), the signals in the tuned and the detuned cases are similar. 
In this time span, the CR oscillation is not affected by the cavity fields and develops as a matter-only excitation.
We developed a quantum electrodynamic model, which allows us to trace in the time domain the occupations of the CR and cavity modes, after excitation of one electron into a higher Landau level (Supplementary Materials Sec.\ VI). In Fig.\ \ref{fig:cross-thick}d shows the vacuum Rabi oscillations of light and matter modes. However, any effect of the cavity on the occupation of states starts only after the 1st cavity cycle, since during that time span the occupation of the CR is identical to the decay of bare 2DEG. We also show the lack of the cavity effect during the first cavity cycle using transfer-matrix-based simulation of the time-domain response in samples of different thicknesses (Supplementary Materials Fig.\ S8).
The effect of the cavity modes, in particular the vacuum Rabi oscillations, emerges only after this delay. This behavior reflects the finite build-up time of the Fabry–Perot cavity field, which is formed by an interference of counter-propagating waves, establishing standing modes. Consequently, the coherent light–matter interaction is not instantaneous but develops on the timescale of the cavity round-trip, directly revealing the dynamical formation of the polaritonic state.

\begin{figure*}
    \centering
    \includegraphics[width=\linewidth]{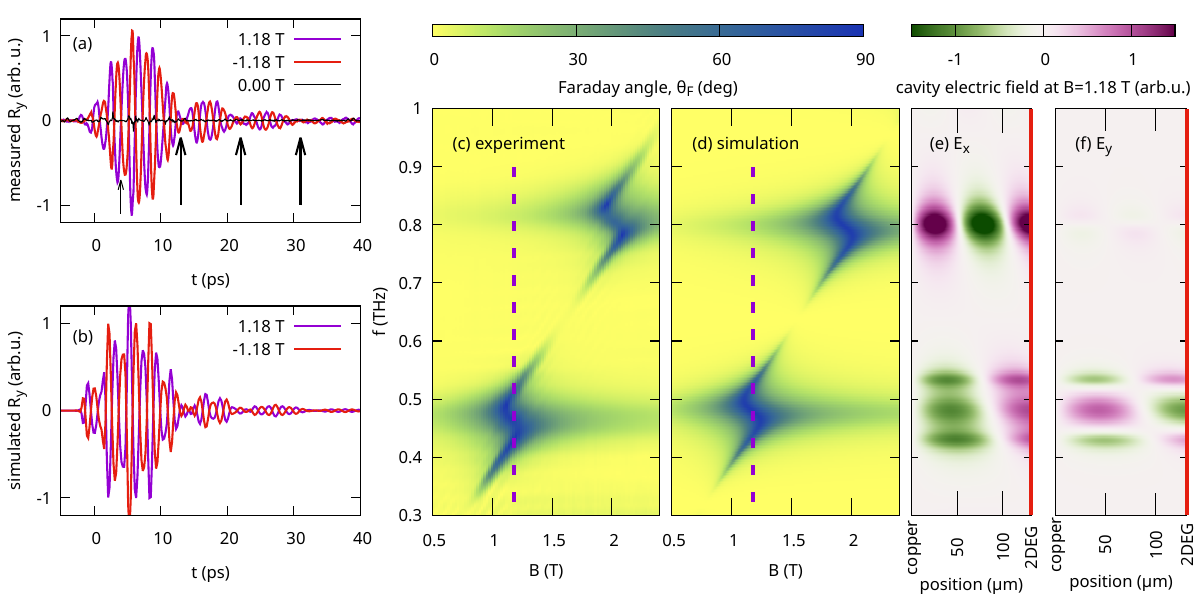}
    \caption{Thin-substrate sample in cross-polarized measurements. (a) Time-domain traces of reflection spectra at $B=0$ T and at $B=\pm1.18$ T. Larger black arrows in the yellow segment mark observed nodes of vacuum Rabi oscillations beating, while the small black arrow in the green segment marks an expected first node. (b) Simulated time-domain traces at $B=1.18$~T.
    (c) Faraday angle spectra $\theta_F$ as a function of magnetic field. Dashed line marks the spectrum at $B=1.18$~T. (d) Simulation of the Faraday angle using the transfer-matrix method.
    Simulation of the cavity mode electric field parallel (e), and perpendicular (f) to the incident field at $B=1.18$~T.}
    \label{fig:cross-thin}
\end{figure*}

In Fig.~\ref{fig:cross-thin}a, we show the cross-polarized time-domain reflection $R_y$ in the case of the thin-substrate sample at $B=1.18$ T, under conditions where the CR frequency is tuned to the cavity mode. The sign of the $R_y$ signal flips with the direction of the magnetic field, and it is negligible at $B=0$ (Supplemental Material Sec.\ S3), unambiguously confirming that it is related to the resonant Faraday effect originating from electrons in the CR.
In contrast, reversal of the CR oscillation with magnetic field sign does not occur in co-polarized $R_x$ time-domain traces, because their phase is defined by the linearly polarized excitation pulse. This demonstrates that the circular nature of the CR response manifests only in the polarization component perpendicular to the polarization of the driving THz field.

The CR oscillations excited by subsequent echoes of the THz pulse (at 0, 3, 6, and 9 ps) dominate the $R_y$ time-domain traces of the thin-substrate sample before 10 ps. These THz pulse echoes are visible directly in the co-polarized traces presented in Supplemental Material Fig.~S3 and discussed in Sec.\ IIB.
After 10 ps, the CR oscillations decay is affected mostly by the strong coupling with the cavity mode, and we observe three distinct minima of the CR oscillation amplitude, marked with black arrows, with a period of 9 ps, significantly longer than the cavity period of 3 ps. This beating pattern in the cross-polarized reflection time-domain trace is a direct demonstration of the periodic energy transfer between light and matter states, that is, the vacuum Rabi oscillations.
In the case of the thin-substrate sample, the vacuum Rabi oscillation period is much shorter than for the thick-substrate sample.

Consequently, a decreased vacuum Rabi oscillation period is reflected in the magnetic-field evolution of the Faraday rotation angle (Fig.\ \ref{fig:cross-thin}c) as larger zigzag patterns related to an increased light-matter coupling, reaching the ultrastrong regime. This results from the reduced volume of the cavity mode $V_m$ and higher electron density $n_s$ of the thin-substrate sample.
Our transfer-matrix modeling reproduces the observed signals (Fig.\ \ref{fig:cross-thin}d).
Based on these fitted frequency-domain spectra, we simulate time-domain traces (Fig.\ \ref{fig:cross-thin}b), which qualitatively reproduce the vacuum Rabi oscillations. Details of the time-domain simulations are described in Supplementary Information Sec.\ 4.3.
In Fig.\ \ref{fig:cross-thin}ef, we show simulated spatial distributions of co- ($E_x$) and cross-polarized ($E_y$) electric fields in the cavity at $B=1.18$~T. In this field, the mode at 0.8 THz does not interact with the CR. This mode is the 3rd Fabry-Perot mode of $5/4\lambda$ profile, allowed by asymmetric boundary conditions of the sample slab placed on the metallic mirror. 
The lower mode at 0.48 THz is the 2nd Fabry-Perot mode of the $3/4\lambda$ profile. This mode interacts strongly with the CR, and three distinct modes are visible in the $E_x$ distribution. Strong coupling with the CR gives the polaritons an electric field component in the cross-polarized direction $E_y$.
The middle mode at 0.48 THz has an opposite sign of $E_y$ to that of the upper and lower Landau-polariton modes, at 0.53 THz and 0.42 THz, respectively. 
This is because the upper and lower polaritons arise from the strong coupling of the co-rotating cavity mode with the CR, while the middle mode is an effect of the weak coupling of the counter-rotating cavity mode with the CR.
We discuss the details of the simulations on light-matter coupling of circularly polarized light in the Supplemental Material Sec.\ S5.

Based on the vacuum Rabi oscillation period as determined in Fig.\ \ref{fig:cross-thin}a, we predict that the first node of the vacuum Rabi oscillations is at about 4 ps, as marked with a small black arrow. This time is after the initial excitation of the CR by the THz pulse at about 0 ps. This indicates that, also in the thin-substrate sample, the light-matter coupling is established after a time offset that is close to the cavity period of 3 ps.

In summary, we demonstrate the strong light–matter coupling in the time domain through phase-resolved measurements, which show vacuum Rabi oscillations. In contrast to conventional frequency-domain power spectroscopy, where only the spectral splitting or amplitude envelope is accessible, our approach directly resolves the coherent energy exchange between light and matter oscillators.
A central result of this work is that the vacuum Rabi oscillations do not emerge immediately after the initial CR excitation, but start with an onset corresponding to one cavity round-trip time. This indicates that the strong-coupling regime is not established instantaneously, but rather dynamically built up through the evolution of the electromagnetic field.
These findings provide direct insight into the temporal formation of hybrid light–matter states and are fundamental for non-equilibrium ultrafast modulation and switching of polaritons. 

\putbib[refs]
\end{bibunit}
\setcounter{section}{0}
\renewcommand{\thesection}{S\arabic{section}}
\renewcommand{\thesubsection}{\thesection.\arabic{subsection}}
\setcounter{figure}{0}
\renewcommand{\thefigure}{S\arabic{figure}}
\begin{bibunit}

 \clearpage

 \setcounter{section}{0}
 \renewcommand{\thesection}{S\arabic{section}}

 \begin{center}
     {\Large\bfseries Supplementary Material}
 \end{center}

\section{Experimental and Samples}
\label{appendix:thick-co}
\begin{figure}
    \centering
    \includegraphics[width=\linewidth]{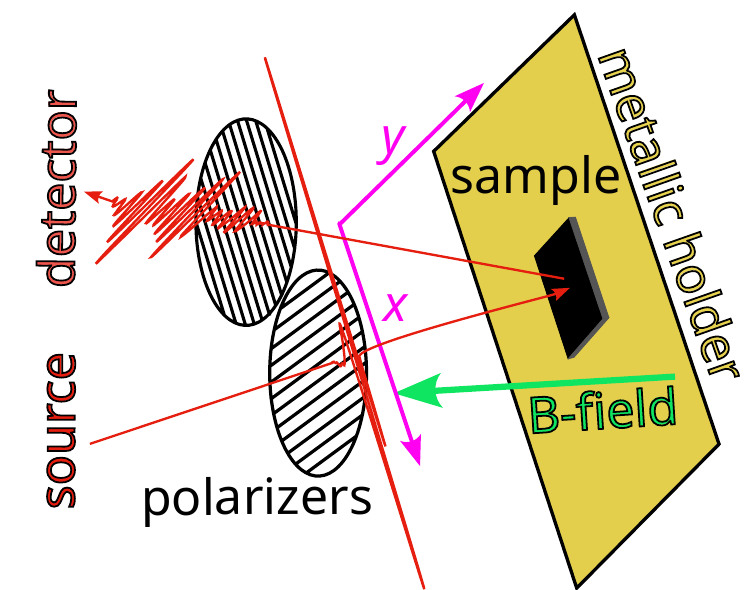}
    \caption{Experimental setup schematics in cross-polarization configuration.}
    \label{fig:setup}
\end{figure}
We use a commercial THz time-domain spectrometer (Toptica) employing photoconductive antennas connected by optical fibers to a pulsed laser operating at 80 MHz repetition rate, 80 fs-long pulses of wavelength centered around 1.5~$\mu$m. The THz reflection path consists of four 1-inch parabolic mirrors, arranged at an incidence angle of approximately 8°. The mirrors that focus the radiation onto the sample and collect the reflected radiation have an effective focal length (EFL) of 4 inches. This configuration provides sufficient working distance to access samples maintained at cryogenic temperatures, while keeping all mirrors at room temperature next to the cryostat windows. The optical setup is mounted on an XYZ stage, facilitating the focusing of THz radiation onto the sample.
To precisely control the THz beam polarization, we use wire-grid polarizers positioned next to the cryostat window. For cross-polarized measurements, we use two wire-grid polarizers that are rotated 90 degrees relative to each other. Such a configuration is possible because the incident and reflected beams are well-separated at the position of the polarizers. Since photoconductive antennas are polarization-sensitive, we adjust the their orientations to match those of the polarizers.

There are two cryostat windows made of z-cut crystalline quartz. The room-temperature window is fixed in the cryostat, while the second window is mounted in the radiation shield. The optical path of the THz beam through each of these windows is approximately 25 ps, which causes the echoes of the THz pulse at about 50 ps after the main pulse. These echoes reexcite the CR in the time domain, leading to additional interference patterns in the frequency domain. Therefore, in this study, we mostly analyze time-domain traces up to 50 ps.
The cryostat operates in a closed-cycle cooling system, allowing temperatures as low as 9 K and magnetic fields of up to 2.5 T. To minimize absorption from atmospheric water vapor, we purge the room-temperature section of the THz beam path with dry nitrogen. The enclosure covering the THz beam path is opaque to visible light, which permits the experiments to be performed in the dark, i.e., without photo-excited carriers.

\begin{table}
    \centering
    \begin{tabular}{lcc}\hline\hline
         & Thick & Thin\\\hline
        Thickness ($\mu$m) & $383\pm2$ & $131\pm2$ \\
        Cavity period (ps) & $9.0\pm0.5$ & $3.0\pm0.5$ \\
        Concentration ($10^{11}$ cm$^{-2}$) & $2.0\pm0.5$ & $4.0\pm0.5$ \\
        Mobility (10$^6$ cm$^2$/Vs) & $>0.5\pm0.2$ & $>0.5\pm0.2$ \\
        Rabi splitting (THz) & $0.04\pm0.01$ & $0.11\pm0.01$ \\
        Rabi beating period (ps) & $22.0\pm0.5$ & $8.5\pm0.5$ \\
        \hline\hline
    \end{tabular}
    \caption{Parameters of the samples and light-matter coupling.}
    \label{tab:samples}
\end{table}
The GaAs/Al$_{0.36}$Ga$_{0.64}$As triangular quantum well is characterized by a high electron mobility of more than $0.5\times10^6$ cm$^2$/Vs and an electron concentration of about $3\times10^{11}$ cm$^{-2}$. The quantum well is about 50 nm under the sample surface. We investigate two samples, each with a lateral dimension of $5\times5$~mm$^2$. The thickness of the first one is the original wafer thickness of $383\,\pm2\,\mu\mathrm{m}$. The second sample has a thickness of $131\,\pm2\,\mu\mathrm{m}$, achieved by mechanical polishing of the wafer substrate. We show the samples' parameters in Tab.\ \ref{tab:samples}. The samples are glued to a copper holder using a thin dielectric layer and placed in the cryostat.

\section{Results in co-polarized reflection}
\subsection{Thick-substrate sample}
\label{appendix:thick-co}
\begin{figure*}
    \centering
    \includegraphics[width=\linewidth]{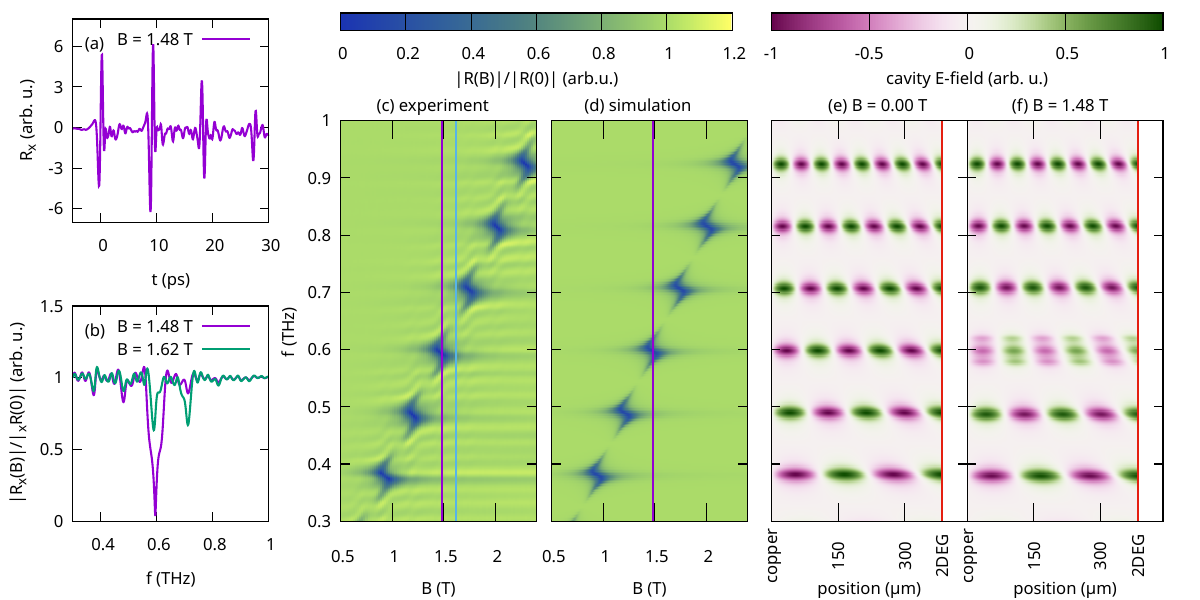}
    \caption{Sample with thick substrate in co-polarized measurements. (a) Time-domain traces of reflection at $B=1.48$ T.
    (b) Normalized frequency-domain spectra at $B=1.48$~T, and $B=1.62$~T.
    (c) Normalized reflection spectra as a function of magnetic field form zigzag patterns due to the strong coupling of the CR with subsequent cavity modes. 
    (d) Simulation of the reflection spectra using the transfer matrix model.
    (e) Simulated distribution of the Fabry-Perot modes at zero magnetic field, (f) and at $B=1.48$~T, shows the three polariton modes of the zigzag pattern near $f = 0.6~\mathrm{THz}$. 
    }
    \label{fig:co-thick}
\end{figure*}
Time-domain co-polarized reflection traces from the thick-substrate sample show multiple peaks related to interface reflections (Fig.\ \ref{fig:co-thick}a), which are separated by about 9 ps, the cavity period, i.e., twice the time-of-flight of the THz pulse in the sample.
The first peak corresponds to the reflection from the sample surface and the 2DEG. In contrast, the second peak originates from the radiation transmitted through the sample and reflected from the copper holder.
The higher amplitude of the second peak results from the nearly unity reflectivity of the dielectric–metal interface.
We note that the first and second peaks have the same phase because both reflections occur at interfaces where the incident radiation propagates from a medium of lower refractive index into a medium of higher refractive index.
In contrast, the subsequent peaks exhibit alternating phase reversals. These arise from successive internal reflections at the dielectric–vacuum interface, where the incident pulse is reflected as it propagates in a medium of higher refractive index than the surrounding medium.

Fourier transforms of these $R_x(t)$ time-domain traces yield $|R_x(f)|$ frequency-domain reflection spectra. In Fig.\ \ref{fig:co-thick}bc, we show such spectra normalized to the spectrum measured at zero magnetic field $|R_x(f,B)|/|R_x(f,B=0)|$.
The reflection spectrum at $B=0$ does not reveal the FP modes, as pulses reflected from the metallic holder interfere constructively.
Under an applied magnetic field, the minima appear in the reflection spectra (Fig.\ \ref{fig:co-thick}c), forming zigzag patterns as a function of $B$, when the CR crosses subsequent Fabry-Perot modes.
The minimum of co-polarized reflectivity drops to near 1\% of the incident electric field. We simulate the reflection spectra (Fig.\ \ref{fig:co-thick}d), using the transfer-matrix method model fitted to the experimental result (for more details, see sec.\ IV). Employing the same approach, we calculate the electric-field distribution at $B = 0~T$ (Fig.\ \ref{fig:co-thick}e)  where only the Fabry-Perot cavity modes are present in the absence of the CR. In Fig.\ \ref{fig:co-thick}f, the electric-field distribution at $B = 1.48~T$, corresponds to the violet line in Fig.\ \ref{fig:co-thick}d. It shows three polariton modes, corresponding to the lower, middle, and upper polariton branches, respectively. The modes in Fig.\ \ref{fig:co-thick}f arise from the coupling between the CR and the $6$-th Fabry-Perot cavity mode at $f_6 = 0.6\, \mathrm{THz}$.

\subsection{Thin-substrate sample}
\label{appendix:thin-co}
\begin{figure*}
    \centering
    \includegraphics[width=\linewidth]{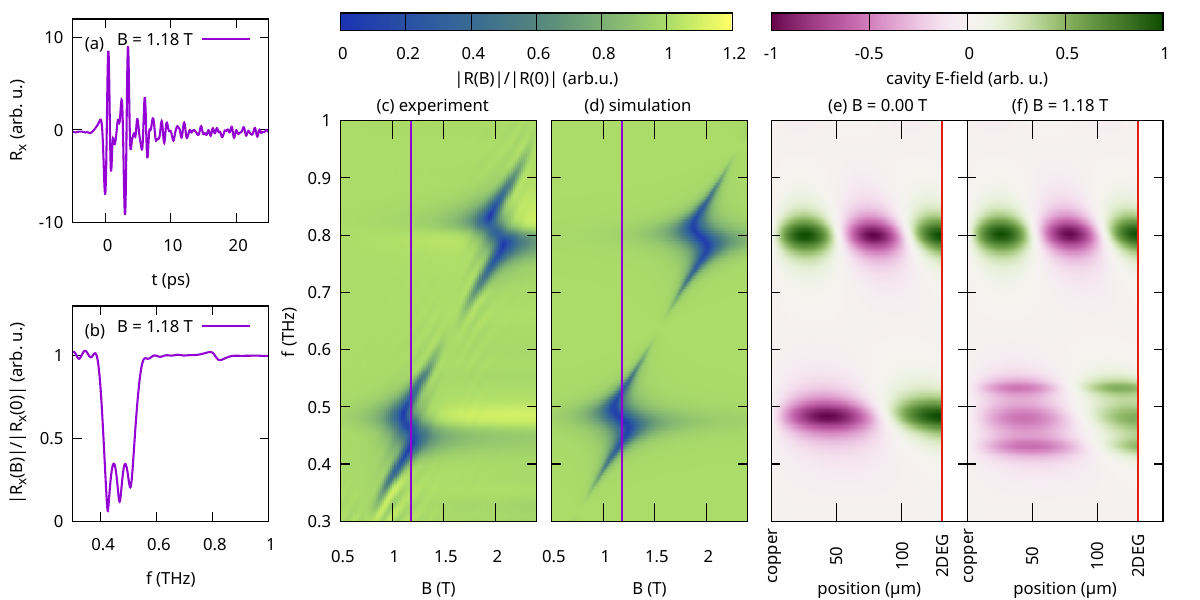}
    \caption{Sample with thin substrate in co-polarized measurements. (a) Time-domain traces of reflection spectra at $B=1.18$ T, and (b) its frequency-domain spectrum. (c) Reflection spectra as a function of magnetic field. The field of $B=1.18$ T is marked with a vertical violet line. (d) Transfer-matrix-based simulation of reflection spectra. (e) Cavity modes electric field profile without magnetic field, and (f) at $B=1.18$~T.}
    \label{fig:co-thin}
\end{figure*}
In the time-domain reflection traces of the thin-substrate sample (Fig.\ \ref{fig:co-thin}a) exhibit reflection peaks associated with the Fabry–Perot cavity that are separated by approximately 3 ps, corresponding to a cavity round-trip time about three times shorter than that of the thick-substrate sample. Consequently, under an applied magnetic field, the CR oscillations are more convoluted with the Fabry–Perot cavity reflection than in the thick-substrate sample.
Since the cavity period in this sample is shorter, there are only two Fabry–Perot modes within the experimentally accessible range of emitter TDS antenna power and magnetic field (Fig.\ \ref{fig:co-thin}b).

\subsection{Delayed formation of Landau polaritons in magnetic-field-differential time traces}
\label{appendi:Bdiff}
\begin{figure*}
    \centering
    \includegraphics[width=\linewidth]{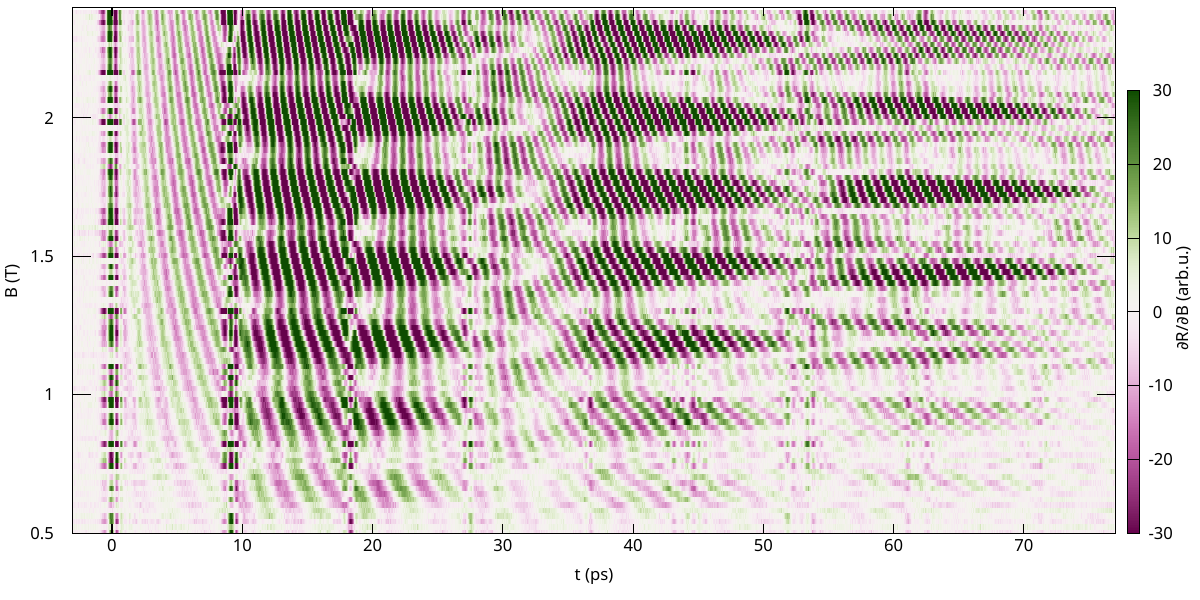}
    \caption{\label{TD-Bdiff}Magnetic-field-differential of time-domain reflection traces of the thick substrate-sample in co-polarization. 
    CR oscillations between the 1st and the 2nd peak (0 to 9 ps) are not affected by the cavity photon field.}
\end{figure*}

The long cavity period observed in the thick-substrate sample allows us to analyze the formation of Landau-polaritons in magnetic-field-differential time-domain traces. 
The first THz peak corresponds to the pulse reflected from the sample surface and the 2DEG, which therefore does not propagate through the substrate.
Between the 1st and 2nd peaks, we observe clear CR oscillations (0 to 0 ps in Fig.\ \ref{TD-Bdiff}), although their amplitude is several times weaker than after the 2nd pulse, which transmitted twice through the 2DEG.

The CR oscillations following the first THz pulse and before the second pulse ($0-9 $~ps in Fig.\ \ref{TD-Bdiff}) show no modulation related to the crossings of the CR frequency with the Fabry-Perot modes. 
The vacuum Rabi oscillation starts only after one full cycle of the cavity has passed. This is reproduced by the quantum model of Sec.~\ref{sec:QED}: during the first cavity round trip, $t<\tau\approx9$~ps, the Kohn-mode occupation decays exactly as for a 2DEG on top of a semi-infinite GaAs slab, and no quantum returns from the field to the electrons (Fig.~\ref{fig:occupations}). 
In the Fabry–Perot cavity, this means the electromagnetic field propagates back and forth within the resonator---which coincides with the arrival of the second THz pulse at the 2DEG. From the present experiment, it is not possible to determine whether the establishment of strong coupling requires the second pulse to re-excite the electrons with a specific phase, or whether it is sufficient for the information about the matter excitation to circulate within the cavity, since both travel with the same speed. However, with the quantum model (Sec.~\ref{sec:QED}), we show that the second mechanism is sufficient. We consider what happens when the Kohn mode is excited by a single impulsive kick at $t=0$~ps. The reversible light-matter exchange begins at $t=\tau$, driven only by the field that the electrons themselves emitted into the substrate one round trip earlier.

The reexcitation of the CR oscillations with pulse echoes produces in the cryostat windows at about 50 ps is still noticeable as an enhancement of the amplitude of the 3rd Rabi beating antinode (at about 60 ps in Fig.\ \ref{TD-Bdiff}a). This increase in amplitude is caused by additional energy delivered to the polaritonic system by the pulse echoes created in the cryostat windows.

\section{Normalization of time-domain traces in cross-polarized reflection}
\begin{figure}
    \centering
    \includegraphics[width=\linewidth]{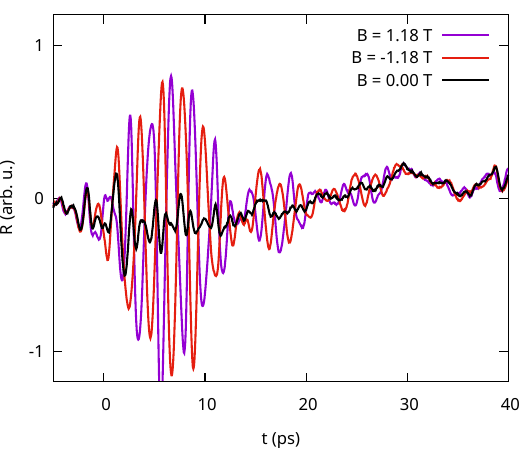}
    \caption{Time-domain reflection traces in cross-polarization at $B = \pm\, 1.18~T$ with slow-varying background}
    \label{fig:LandR1}
\end{figure}
Time-domain cross-polarized reflection measurements (Fig.\ \ref{fig:LandR1}) exhibit strongly suppressed peaks at zero magnetic field.
However, these time-domain traces contain a slowly varying background, which generates spectral weight only below 0.2 THz and is therefore negligible in the relevant frequency range.
We subtract this background to better focus attention on fast oscillations caused by the CR.
Specifically, we subtract the median of all traces measured between $-2.5$ and $+2.5$ T. This background trace is similar, but not identical, to the zero-field trace.

\section{Transfer matrix model}
\label{TMM}
To provide a theoretical description that accounts for the 2DEG conductivity, the underlying semi-insulating GaAs substrate, and the metallic holder, we employ the transfer matrix method (TMM), a standard approach for modeling optical and electromagnetic wave propagation \cite{BornWolf}. We consider a plane electromagnetic wave incident normally from vacuum onto the structure.
We approximate the incident angle of $8^{\circ}$ as normal incidence in the simulations, since its effect on the optical response is negligible. The 2DEG is assumed to be located at the surface of a GaAs substrate layer with thickness $d$ and refractive index $n= \sqrt\varepsilon= 3.6$ backed by a semi-infinite copper (Cu) layer (Fig. \ref{Simplified structure}). 

\begin{figure}[h]
\centering
\begin{tikzpicture}[scale=1]

\fill[gray!20] (-1.0,0) rectangle (3.5,-1);
\fill[yellow!80] (-1.0,-1) rectangle (3.5,-3);

\draw[red, line width=1.2pt] (-1.0,0) -- (3.5,0);

\draw[dashed, thin, blue!60] (-3.5,-0.03) -- (-1.0,-0.03);
\draw[dashed, thin, blue!60] (-3.5,-1) -- (-1.0,-1);

\draw[->, very thick, blue!70] (-4.0,0.8) -- (-4.0,-3.2);
\node[below] at (-4.4,-2.9) {$z$};

\node[right] at (-4.7,0) {$z=0$};
\node[right] at (-4.7,-1) {$z=d$};

\node at (1.2,-0.5) {GaAs ($\varepsilon$)};
\node at (1.2,-2) {Cu (semi-infinite)};
\node[above] at (1.2,-0.1) {2DEG ($\sigma_{\pm}$)};

\draw[-{Latex}, thin] (-3,0.9) -- (-3,0.1);
\node[left] at (-3,0.6) {$E_0^{\rightarrow\pm}$};

\draw[-{Latex}, thin] (-1.5,0.1) -- (-1.5,0.9);
\node[left] at (-1.5,0.6) {$E_0^{\leftarrow\pm}$};

\draw[-{Latex}, thin] (-3,-0.1) -- (-3,-0.9);
\node[left] at (-3,-0.3) {$E_1^{\rightarrow\pm}$};

\draw[-{Latex}, thin] (-1.5,-0.9) -- (-1.5,-0.1);
\node[left] at (-1.5,-0.3) {$E_1^{\leftarrow\pm}$};

\draw[-{Latex}, thin] (-3,-1.1) -- (-3,-1.9);
\node[left] at (-3,-1.3) {$E_2^{\rightarrow\pm}$};

\end{tikzpicture}
\hfill
\caption{Geometry of the vacuum/2DEG/GaAs/Cu multilayer structure considered for reflection calculations under a perpendicular magnetic field.}
\label{Simplified structure}
\end{figure}

\subsection{Reflection Coefficients}
Since the experimental measurements are performed using linearly polarized THz radiation, the incident field is decomposed into right- and left-circularly polarized (RCP and LCP) components, which are treated independently and subsequently recombined to recover the linearly polarized response. The propagation of electromagnetic waves through the multilayer system is therefore calculated for each circular polarization, the electric field inside the GaAs layer is represented as a superposition of forward- and backward-propagating plane waves.
\begin{equation}
E_1^{\pm}(z) = E_1^{\rightarrow\pm}\, e^{iqz}\, + \, E_1^{\leftarrow\pm}\, e^{-i qz},
\end{equation}

where $q = \frac{\omega}{c}\sqrt{\varepsilon}$ corresponds to the z-component of the wave vector inside the layer. The coefficients \(E_1^{\rightarrow\pm}\) and \(E_1^{\leftarrow\pm}\) denote the amplitudes of the forward- and backward-propagating waves, respectively, while the superscripts \((+)\) and \((-)\) correspond to the right- and left-circularly polarized (RCP and LCP) components. While the propagation through the GaAs substrate is governed by its dielectric response, the magneto-optical response of the quantum well is described through an effective surface (sheet) conductivity. Since the quantum well barrier is approximately $50$-nm-thick, the 2DEG can be treated as an infinitesimally thin conducting sheet \cite{Dzhikirba2024}. In the presence of a perpendicular magnetic field, this conductivity becomes tonsorial. Transforming to the circular polarization basis diagonalizes the conductivity tensor, yielding the following expression for the two-dimensional conductivity:

\begin{equation}
\sigma_{\pm} = \dfrac{\sigma_{DC}}{\ 1 - i(\omega\;{\mp }\;\omega_{c})\tau}, 
\quad   
\sigma_{DC} = \dfrac{n_{s}e^{2}\tau}{m^{*}},
\quad
\omega_{c} = \dfrac{eB}{m^{*}}
\end{equation}

Where $\sigma_{\pm}$ are the 2DEG conductivities for RCP \((+)\) and LCP \((-)\) components, respectively. They are determined by electron concentration $n_{s}$, electron scattering time $\tau = \dfrac{\mu m^{*}}{e}$, and electron effective mass $m^{*} = 0.067\;m_{e}$. Samples-specific parameters are given in Tab.~\ref{tab:samples}.

For each circular polarization, the tangential electric and magnetic fields at the entrance ($z=0$) and exit ($z=d$) of the structure are related through the total transfer matrix by,

\begin{equation}
\begin{pmatrix}
E_0^{\pm} \\\\[8pt]
H_0^{\pm}
\end{pmatrix}
=
Q_{ij}^{\pm}
\begin{pmatrix}
E_d^{\pm} \\\\[8pt]
H_d^{\pm}
\label{Total Transfer Matrix}
\end{pmatrix}
\end{equation}
\quad 

Where, $Q_{ij}^{\pm} = M_{2\mathrm{DEG}}^{\pm}\,
M_{\mathrm{GaAs}}(d)$,  with $i,j = 1,2$

\begin{equation}
M_{2\mathrm{DEG}}^{\pm}
=
\begin{pmatrix}
1 & 0 \\ \\[8pt]
Z_0 \sigma_{\pm} & 1
\end{pmatrix}
\end{equation}

\begin{equation}
M_{\mathrm{GaAs}}(d)
=
\begin{pmatrix}
\cos(qd) & -\dfrac{i}{\sqrt{\varepsilon}}\sin(qd) \\\\[8pt]
-i\sqrt{\varepsilon}\sin(qd) & \cos(qd)
\end{pmatrix}.
\end{equation}
 
Here, $Z_{0} = \sqrt{\dfrac{\mu_0}{\varepsilon_{0}}}$ denotes the impedance of free space, and the matrices $M_{2\mathrm{DEG}}^{\pm}$ describe the response of the 2DEG for RCP and LCP, respectively. They are obtained from the continuity of the tangential electric field and the discontinuity of the tangential magnetic field across the conducting 2DEG sheet.

The semi-infinite Cu layer is modeled using the Drude dielectric function
\begin{equation}  
\varepsilon_{\mathrm{Cu}}(\omega) = \varepsilon_{\infty} - \frac{\omega_{p}^{2}}{\omega(\omega + i\gamma)},
\end{equation}
where $\varepsilon_{\infty} = 1$, $\omega_p/2\pi \approx 2.61 \times 10^{4}\,\mathrm{THz}$, and $\gamma/2\pi \approx 22 \,\mathrm{THz}$ are the high-frequency dielectric constant, plasma frequency, and damping rate of copper, respectively \cite{Johnson1972}. In the THz frequency range ($\omega \ll \gamma \ll \omega_{p}$), this model yields a large negative permittivity dominated by its imaginary part. Consequently, copper behaves effectively as a near-perfect electric conductor.

\quad

Given the incident electric and magnetic field amplitudes $E_0^{\pm} = E_0^{\rightarrow\pm} + E_0^{\leftarrow\pm} $, $H_{\mathrm{0}} = n_{air}(E_0^{\rightarrow} - E_0^{\leftarrow}$), respectively, by assuming that no backward waves are propagating in the copper  ($E_d^{\pm} = E_2^{\rightarrow\pm}$, $H_d^{\pm} = \sqrt{\varepsilon_{Cu}}E_2^{\rightarrow\pm}$). The reflected fields in each layer are obtained with electromagnetic boundary conditions at each interface. The multilayer system leads to the following matrix equation:

\begin{equation}
\begin{pmatrix}
E_0^{\rightarrow\pm} + E_0^{\leftarrow\pm}  \\\\[8pt]
n_{vac}\;(E_0^{\rightarrow\pm} - E_0^{\leftarrow\pm})
\end{pmatrix}
=
Q_{ij}^{\pm}
\begin{pmatrix}
E_2^{\rightarrow\pm} \\\\[8pt]
\sqrt{\varepsilon_{Cu}}\,E_2^{\rightarrow\pm}
\label{Total Transfer Matrix}
\end{pmatrix}
\end{equation}

The reflection coefficients for RCP and LCP light are calculated independently from the corresponding total transfer matrices (Eq. \ref{Total Transfer Matrix}). They are defined as the ratio of the backward- to forward-propagating electric field amplitudes at the incident side, $E_0^{\leftarrow\pm} / E_0^{\rightarrow\pm}$, yielding :

\begin{equation}
R_{\pm} = \dfrac{( Q_{11}^{\pm} + \sqrt{\varepsilon_{Cu}}\,Q_{12}^{\pm} ) - ( Q_{21}^{\pm} + \sqrt{\varepsilon_{Cu}}\,Q_{22}^{\pm} )}{( Q_{11}^{\pm} + \sqrt{\varepsilon_{Cu}}\,Q_{12}^{\pm} ) + ( Q_{21}^{\pm} +\sqrt{\varepsilon_{Cu}}\,Q_{22}^{\pm} )}
\label{RCP and LCP reflection}
\end{equation}
Here, $Q_{ij}^{\pm}$ denotes the elements of the total transfer matrix for RCP and LCP light, respectively. The corresponding reflection coefficients for the two circular polarizations are defined as $R_{+}$ and $R_{-}$, respectively.

The linear reflected field is therefore obtained by recombining these two circular contributions. The co-polarized  (parallel) reflection coefficient $R_{x}$ is given by:
\begin{equation}
R_{x} = \frac{R_{+} + R_{-}}{2}.
\end{equation}
The cross-polarized (orthogonal) reflection coefficient $R_{y}$ is obtained from
\begin{equation}
R_{y} = \frac{R_{+} - R_{-}}{2i}.
\end{equation}

\begin{figure}[h]
\centering
\begin{tikzpicture}[scale=1.4]

\def\w{1.45}
\def\eps{0.05}


\fill[white] (0.2,0) rectangle ++(\w,2);
\node[above,font=\small] at (0.2+0.8*\w,2.1) {Air};

\fill[red!60] (0.2+\w,0) rectangle ++(\w,2);
\node[above,font=\small] at (0.2+1.5*\w,2.1) {2DEG};

\fill[gray!20] (0.2+2*\w,0) rectangle ++(\w,2);
\node[above,font=\small] at (0.2+2.5*\w,2.1) {GaAs};

\fill[yellow!80] (0.2+3*\w,0) rectangle ++(\w,2);
\node[above,font=\small] at (0.2+3.5*\w,2.1) {Cu};



\draw[-{Latex[length=3.5pt,width=2.5pt]},thin]
(0.35,1.5) -- (1.45,1.5)
node[midway,above,font=\footnotesize]
{$E_{0}^{\rightarrow\pm}$};

\draw[-{Latex[length=3.5pt,width=2.5pt]},thin]
(1.5,0.5) -- (0.35,0.5)
node[midway,below,font=\footnotesize]
{$E_{0}^{\leftarrow\pm}$};

\def\x{0.2+\w}

\draw[-{Latex[length=3.5pt,width=2.5pt]},thin]
(\x+\eps,1.5) -- (\x+0.9,1.5)
node[midway,above,font=\footnotesize]
{$E_{1}^{\rightarrow\pm}$};

\draw[-{Latex[length=3.5pt,width=2.5pt]},thin]
(\x+0.9,0.5) -- (\x+\eps,0.5)
node[midway,below,font=\footnotesize]
{$E_{1}^{\leftarrow\pm}$};

\def\x{0.2+2*\w}

\draw[-{Latex[length=3.5pt,width=2.5pt]},thin]
(\x+\eps,1.5) -- (\x+0.9,1.5)
node[midway,above,font=\footnotesize]
{$E_{2}^{\rightarrow\pm}$};

\draw[-{Latex[length=3.5pt,width=2.5pt]},thin]
(\x+0.9,0.5) -- (\x+\eps,0.5)
node[midway,below,font=\footnotesize]
{$E_{2}^{\leftarrow\pm}$};

\def\x{0.2+3*\w}

\draw[-{Latex[length=3.5pt,width=2.5pt]},thin]
(\x+\eps,1.5) -- (\x+1.0,1.5)
node[midway,above,font=\footnotesize]
{$E_{3}^{\rightarrow\pm}$};


\node[font=\small,text=blue!70]
at (0.2+1*\w,1)
{$I_{0,1}^{\pm}$};

\node[font=\small,text=blue!70]
at (0.2+2*\w,1)
{$I_{1,2}^{\pm}$};

\node[font=\small,text=blue!70]
at (0.2+3*\w,1)
{$I_{2,3}^{\pm}$};


\node at (0.2+1.5*\w,1)
{$P_{1}^{\pm}$};

\node at (0.2+2.5*\w,1)
{$P_{2}^{\pm}$};


\draw[dashed,red!70] (0.2+\w,0) -- (0.2+\w,-1.2);
\draw[dashed,red!70] (0.2+2*\w,0) -- (0.2+2*\w,-1.2);
\draw[dashed,gray!70] (0.2+3*\w,0) -- (0.2+3*\w,-1.2);


\node[below] at (0.6,0)
{\small Layers:};

\node[below] at (0.2+0.8*\w,0)
{\small 0};

\node[below] at (0.2+1.5*\w,0)
{\small 1};

\node[below] at (0.2+2.5*\w,0)
{\small 2};

\node[below] at (0.2+3.5*\w,0)
{\small 3};


\node[below] at (0.6,-0.4)
{\small Thickness:};

\node[below] at (0.2+0.8*\w,-0.45)
{\small $\infty$};

\node[below] at (0.2+1.5*\w,-0.4)
{\small $50\,\mathrm{nm}$};

\node[below] at (0.2+2.5*\w,-0.4)
{\small $383\,\mu\mathrm{m}$ / $131\,\mu\mathrm{m}$};

\node[below] at (0.2+3.5*\w,-0.4)
{\small Semi-$\infty$};


\node[below] at (0.6,-0.8)
{\small R.I.:};

\node[below] at (0.2+0.8*\w,-0.8)
{\small $1$};

\node[below] at (0.2+1.5*\w,-0.8)
{\small $\sqrt{\varepsilon_{2DEG}^{\pm}}$};

\node[below] at (0.2+2.5*\w,-0.8)
{\small $\sqrt{\varepsilon_{GaAs}}$};

\node[below] at (0.2+3.5*\w,-0.8)
{\small $\sqrt{\varepsilon_{Cu}}$};

\end{tikzpicture}

\caption{Transfer-matrix representation of the Air/2DEG/GaAs/Cu structure showing the forward- and backward-propagating field amplitudes, interface matrices $I_{N,N+1}^{\pm}$, and propagation matrices $P_N^{\pm}$ for the RCP and LCP.}
\label{fig:TMM_structure}
\end{figure}

\subsection{Electric field distribution}
The electromagnetic field distribution inside the multilayer structure is obtained using TMM. The field amplitudes are propagated through successive interfaces and layers via interface ($I$) and propagation ($P$) matrices:

\begin{equation}
\mathbf{V}_{N+1} = P_N \, I_{N,N+1} \, \mathbf{V}_N,
\qquad
\mathbf{V}_0 =
\begin{pmatrix}
1 \\
R_{\pm}
\end{pmatrix},
\end{equation}

Where;

\begin{equation}
I_{N,N+1}^{\pm}
=
\frac{1}{2}
\begin{pmatrix}
1+\dfrac{n_N^{\pm}}{n_{N+1}^{\pm}}
&
1-\dfrac{n_N^{\pm}}{n_{N+1}^{\pm}}
\\\\[8pt]
1-\dfrac{n_N^{\pm}}{n_{N+1}^{\pm}}
&
1+\dfrac{n_N^{\pm}}{n_{N+1}^{\pm}}
\end{pmatrix};
\quad N = 0,1,2.
\end{equation}

\begin{equation}
P_N^{\pm}
=
\begin{pmatrix}
e^{iq_N^{\pm}d_N}
&
0
\\\\[8pt]
0
&
e^{-iq_N^{\pm}d_N}
\end{pmatrix};
\quad N = 1,2.
\end{equation}

where $\mathbf{V}_N = (V^{\pm}_1, V^{\pm}_2)^T$ represents the forward and backward propagating field amplitudes in each layer.

The electric field inside a given layer is expressed as:

\begin{equation}
E^{\pm}(z) = V^{\pm}_1\, e^{i q^{\pm} z} + V^{\pm}_2\, e^{-i q^{\pm} z},
\end{equation}
where $(+)$ and $(-)$ correspond to RCP and LCP, respectively.

The measurable linear polarization components are reconstructed from the circular basis as:

\begin{equation}
E_x(z) = \frac{E^{+}(z) + E^{-}(z)}{2},
\qquad
E_y(z) = \frac{E^{+}(z) - E^{-}(z)}{2i}.
\end{equation}

This decomposition allows direct evaluation of the magneto-optical response of the structure in the linear polarization basis. In particular, it provides a practical framework for analyzing the coupling between FP cavity resonances and the CR of 2DEG, which manifests itself as polarization conversion and mode hybridization in the reflection spectra.

\subsection{Time-domain simulations}
To obtain the time-domain response, the calculated complex reflection coefficients for the co- and cross-polarized reflected fields were multiplied by the spectrum of a Gaussian THz pulse centered at $f_0 = 1.0~\mathrm{THz}$ with a spectral width of $\sigma_f = 0.5~\mathrm{THz}$ and a constant phase offset $\phi = \dfrac{\pi}{2}$.

\begin{equation}
    E_{\mathrm{in}}(f) = \exp\left[{-\dfrac{(f - f_0)^2}{2\sigma_{f}^2}}\right]e^{i\phi},
\end{equation}

The reflected spectra for the co- and cross-polarized components were calculated as

\begin{align}
E_{\mathrm{ref},x}(f) &= R_x(f)\,E_{\mathrm{in}}(f) ; \qquad
E_{\mathrm{ref},y}(f) &= R_y(f)\,E_{\mathrm{in}}(f).
\end{align}

Where $R_x(f)$ and $R_y(f)$ are the complex reflection coefficients obtained from the transfer-matrix calculations.   

The reflected spectra were transformed into the time domain by applying an inverse fast Fourier transform (IFFT). Before performing the transformation, the calculated positive-frequency spectrum was extended to negative frequency by imposing Hermitian symmetry, 
\begin{equation}
    E(-f) = E^*(f),
\end{equation}
Ensuring that the resulting temporal electric field is purely real. The temporal signal was then obtained as 
\begin{equation}
    E_{ref,\alpha} (t)= \mathcal{F}^{-1}\left\{{E_{ref,\alpha}(f)}\right\}; \qquad \alpha = x,y.
\end{equation}
Where $\mathcal{F}^{-1}$ denotes the inverse Fourier transform. The frequency spacing determines the temporal sampling interval according to,

\begin{equation}
    \Delta t = \frac{1}{N\,\Delta f}
\end{equation}
where $N$ is the total number of frequency points after symmetrization and $\Delta f$ is the frequency resolution.

\begin{figure}
    \centering
    \includegraphics[width=\linewidth]{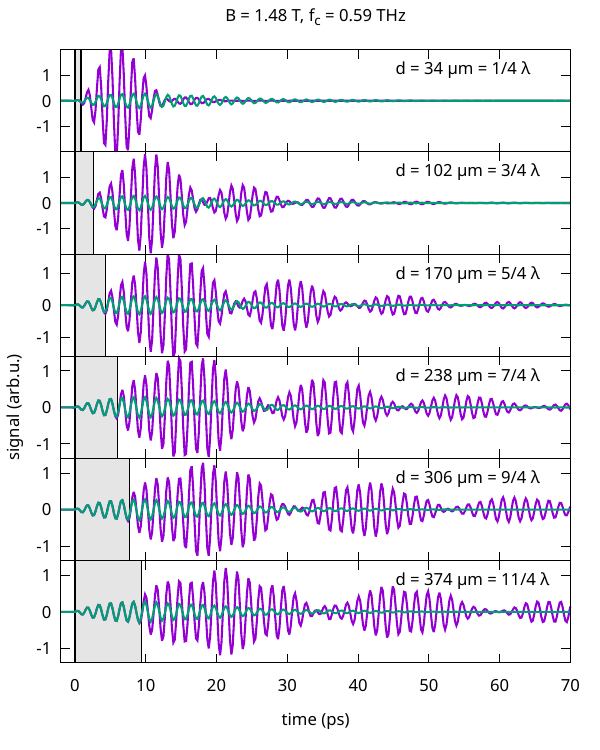}
    \caption{Transfer-matrix-based simulations of the time-domain signal (violet lines) for GaAs substrate thicknesses $d$ that form Fabry-Perot modes resonant with the CR at $B=1.48$~T. The area in gray rectangles corresponds to the cavity period. In this time span, the signal is identical to that of about a 35-mm-thick sample (green curves).}
    \label{fig:TD-by-TMM}
\end{figure}

Using time-domain simulation, we show that Landau-polaritons only start after one cavity cycle. In Fig.\ \ref{fig:TD-by-TMM}, we show simulated time-domain traces for samples of different thicknesses, which form subsequent Fabry-Perot modes resonant with the CR at $B=1.48$~T of $f_c=0.59$~THz.
The response does not depend on the cavity for $t<\tau$. The simulated traces at each resonant thickness match those of a 2DEG on a 35-mm-thick substrate (Fig.~\ref{fig:TD-by-TMM}), and the measured $B$-differential CR oscillations do not depend on tuning to the cavity modes (Fig.~\ref{TD-Bdiff}). After $\tau$, the envelope rises by about an order of magnitude in the tuned cases and drops in the detuned ones. A signal carried by polaritons, which contain cavity amplitude, must depend on $d$ and on the mirror.

\section{Simulation of LCP and RCP reflection}
\label{appendix:LandR}
\begin{figure}
    \centering
    \includegraphics[width=\linewidth]{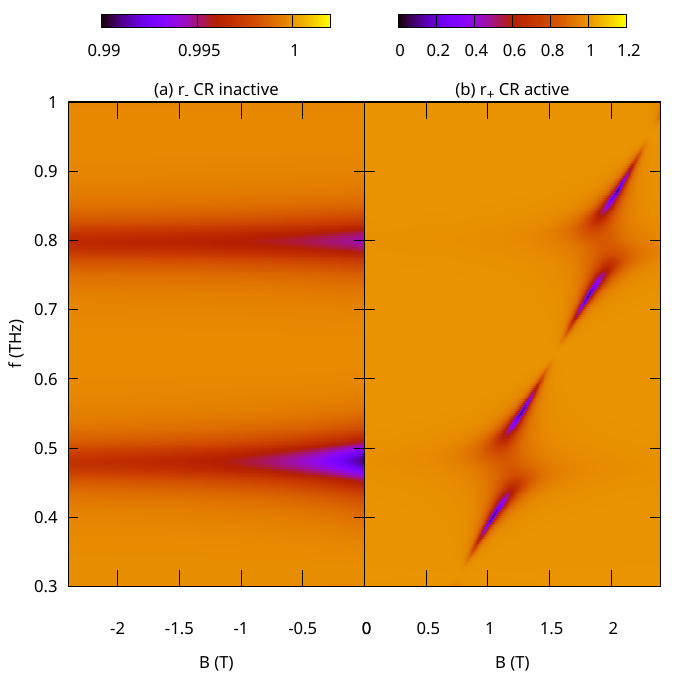}
    \caption{Simulated reflection of an electromagnetic wave from the sample with a thin substrate in the circular polarization basis: (a) LCP (CR inactive) and (b) RCP (CR active).}
    \label{fig:LandR}
\end{figure}
Our simulations (Fig.~\ref{fig:LandR}) show that for RCP light, which couples to the cyclotron resonance, clear anticrossings appear, corresponding to the formation of the upper and lower Landau polariton branches. 

In contrast, for LCP, the CR is inactive and only weak light–matter interaction remains. As a result, the Fabry–Perot cavity modes exhibit only a very small magnetic-field dependence, visible mainly as a slight change in their reflection amplitude less than $1\%$ for LCP in Fig.~\ref{fig:LandR}a. Thus, a simple summation of the amplitudes of the LCP and LCR, as presented in Fig.\ \ref{fig:LandR}, does not predict the formation of the zigzag pattern. It originates from the phase of the circular reflection coefficients $r_-$ and $r_+$. The characteristic zigzag patterns are reproduced only when we sum complex reflection coefficients related to the two circular polarizations. Thus, it is the phase of these two reflection spectra that is producing the zigzag patterns, as well as the related excitation of CR with counter-rotating circular polarization, which can be interpreted as the vacuum Broch-Siegert effect \cite{Li18}.

\section{Quantum model of delayed Landau-polariton formation}
\label{sec:QED}
The transfer-matrix model of Sec.~\ref{TMM} reproduces the reflection spectra in the time- and frequency domains, and allows us to simulate results in samples of different geometries. However, the occupations of light and matter states have to be resolved using a quantum model of the 2DEG coupled to the field of the open substrate cavity. The calculated occupations undoubtedly show the delayed onset of the periodic light-matter energy exchange related to the vacuum Rabi oscillations.

\subsection{Hamiltonian and retarded equations of motion}
\label{sec:QED-H}
By Kohn's theorem~\cite{Kohn1961}, the cyclotron resonance of a translationally invariant 2DEG is a single collective harmonic mode, the Kohn mode, with bosonic operator $b$ ($[b,b^\dagger]=1$) and frequency $\omega_c=eB/m^*$. Only the CR-active circular component couples to it, and we treat that component in the rotating-wave approximation. The 2DEG sits at $z=0$, at the vacuum/GaAs interface, and radiates into two field continua: modes $a_\omega$ propagating in vacuum and modes $c_\omega$ propagating in the substrate. The Hamiltonian is
\begin{align}
H ={}& \hbar\omega_c\, b^\dagger b
+ \int\! d\omega\, \hbar\omega \left[a^\dagger_\omega a_\omega + c^\dagger_\omega c_\omega\right] \nonumber\\
&+ i\hbar \int\! \frac{d\omega}{\sqrt{2\pi}} \Big\{ \sqrt{\gamma_1}\left[a^\dagger_\omega b - b^\dagger a_\omega\right] \nonumber\\
&\qquad\qquad + \sqrt{\gamma_2}\left[c^\dagger_\omega b - b^\dagger c_\omega\right]\Big\}
+ H_{\mathrm{nr}},
\label{eq:QED-H}
\end{align}
where $H_{\mathrm{nr}}$ couples $b$ to the electronic bath responsible for the non-radiative damping $\gamma_{\mathrm{nr}}$. The radiative rates into vacuum ($\gamma_1$) and into the substrate ($\gamma_2$) are fixed by the electron density,
\begin{equation}
\gamma_{\mathrm{rad}} = \gamma_1+\gamma_2 = \frac{2 n_s e^2}{m^* \varepsilon_0 c\,(1+n)},
\label{eq:QED-rates}
\end{equation}
with $\gamma_1=\gamma_{\mathrm{rad}}/(1+n)$, $\gamma_2=n\gamma_{\mathrm{rad}}/(1+n)$, and $n=\sqrt{\varepsilon}$ the refractive index of GaAs. The collective coupling scales as $\sqrt{n_s}$, and $\gamma_{\mathrm{rad}}$ is the \textit{superradiant} decay rate of the CR of a 2DEG on a GaAs half-space~\cite{Zhang2014}. The copper holder at $z=d$ closes the substrate into a Fabry-Perot cavity. Eliminating the field continua in the standard input-output manner~\cite{Gardiner1985} gives the Heisenberg-Langevin equations
\begin{align}
\dot b(t) &= -\left(i\omega_c+\tfrac{\gamma}{2}\right) b(t) + \sqrt{\gamma_1}\, a_{\mathrm{in}}(t) \nonumber\\
&\quad + \sqrt{\gamma_2}\, c_{\mathrm{in}}(t) + F_{\mathrm{nr}}(t),
\label{eq:QED-b}\\
c_{\mathrm{out}}(t) &= t'\, a_{\mathrm{in}}(t) + r\, c_{\mathrm{in}}(t) - \sqrt{\gamma_2}\, b(t),
\label{eq:QED-cout}\\
c_{\mathrm{in}}(t) &= r_m\, c_{\mathrm{out}}(t-\tau),
\label{eq:QED-delay}\\
a_{\mathrm{out}}(t) &= -r\, a_{\mathrm{in}}(t) + t' c_{\mathrm{in}}(t) - \sqrt{\gamma_1}\, b(t),
\label{eq:QED-aout}
\end{align}
where $\gamma=\gamma_{\mathrm{nr}}+\gamma_{\mathrm{rad}}$, $F_{\mathrm{nr}}$ is the noise of the non-radiative bath, $r=(n-1)/(n+1)$ and $t'=2\sqrt{n}/(n+1)$ are the internal reflection and transmission amplitudes of the GaAs/vacuum interface, and $r_m$ is the amplitude reflectivity of the metallic mirror. The fields are normalised so that $\langle c^\dagger c\rangle$ is a photon flux. Equation~\eqref{eq:QED-delay} carries the essential physics: the field the electrons emit into the substrate returns to them only after one round trip $\tau$. The 2DEG is coupled to the radiation field from $t=0$, but the cavity feedback, which makes the exchange reversible, is retarded by $\tau$. Equivalently, $b$ couples to the whole continuum of modes of the open slab, in which the Fabry-Perot modes appear as peaks of the density of states; a single cavity mode, with spacing $2\pi/\tau$ to its neighbours, cannot be resolved in a time shorter than $\tau$.

Since $H$ is quadratic, Eqs.~\eqref{eq:QED-b}--\eqref{eq:QED-aout} are linear. The mean amplitude $\beta(t)=\langle b(t)\rangle$ obeys the same equations without the noise term. For a coherent or single-quantum initial excitation of the Kohn mode, with the field initially in vacuum, the occupations are given exactly by $|\beta(t)|^2/|\beta_0|^2$ and by the corresponding squared field amplitudes, per initial quantum.

\subsection{Stationary limit and parameters with CR reexcitations}
\label{sec:QED-linear}
\begin{figure}
     \centering
     \includegraphics[width=\linewidth]{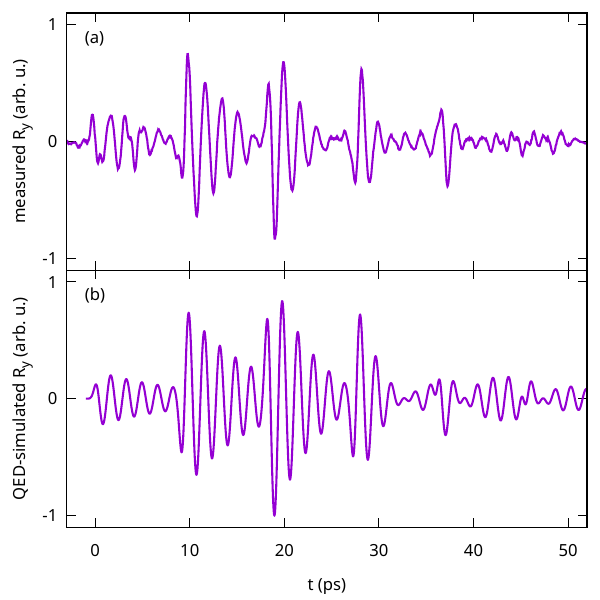}
\caption{Comparison of cross-polarized reflection $R_y$ at $B=1.45$~T. (a) experiment and (b) simulation based on the QED model including subsequent reexcitations of the CR by echoes of the THz pulse.}
     \label{fig:QED-linear}
\end{figure}
For a monochromatic input $a_{\mathrm{in}}\propto e^{-i\omega t}$, Eqs.~\eqref{eq:QED-b}--\eqref{eq:QED-aout} give the reflection coefficient $a_{\mathrm{out}}/a_{\mathrm{in}}$ of the vacuum/2DEG/GaAs/mirror stack. It coincides with Eq.~\eqref{RCP and LCP reflection} for the CR-active component (Fig.\ \ref{fig:LandR}b), with the Drude conductivity $\sigma_+$ in the rotating-wave approximation, $\gamma_{\mathrm{nr}}=2/\tau_{\mathrm{tr}}$ (where $\tau_{\mathrm{tr}}$ is the scattering time denoted $\tau$ in Sec.~\ref{TMM}), and the copper replaced by a mirror of reflectivity $r_m$. We use $n=3.55$, $m^*=0.067\,m_e$, $r_m=-0.99$ and $\gamma_{\mathrm{nr}}=0.05$~ps$^{-1}$ (corresponding to an electron mobility of about $10^6$~cm$^2$/Vs). The electron densities are fixed by requiring the Rabi periods of Tab.~\ref{tab:samples}. They lie within the measured ranges. The resulting parameters are listed in Tab.~\ref{tab:QED}. We solve the time-domain equations by the method of steps.

\begin{table}
    \centering
    \begin{tabular}{lcc}\hline\hline
         & Thick & Thin\\\hline
        $d$ ($\mu$m) & 383 & 131 \\
        $n_s$ ($10^{11}$ cm$^{-2}$) & 2.35 & 4.74 \\
        cavity period $\tau=2nd/c$ (ps) & 9.07 & 3.10 \\
        FP mode $m$; $f_m=\frac{2m-1}{2\tau}$ (THz) & 6; 0.606 & 2; 0.483 \\
        Resonant field $B$ (T) & 1.45 & 1.16 \\
        $\gamma_{\mathrm{rad}}/2$ (ps$^{-1}$) & 0.082 & 0.165 \\
        $\Gamma_0$ (ps$^{-1}$) & 0.107 & 0.190 \\
        Polariton frequencies (THz) & 0.584, 0.629 & 0.428, 0.539 \\
        Polariton linewidth (ps$^{-1}$) & 0.035 & 0.088 \\
        Rabi period $T_R$ (ps) & 22.0 & 9.0 \\
        $\Omega_R\tau/2\pi=\tau/T_R$ & 0.41 & 0.34 \\
        \hline\hline
    \end{tabular}
    \caption{Parameters of the quantum model. The polariton frequencies and linewidths are the two roots of Eq.~\eqref{eq:QED-char} closest to the resonant Fabry-Perot mode.}
    \label{tab:QED}
\end{table}

\subsection{Exact solution without CR reexcitations}
\label{sec:QED-kick}
We consider the Kohn mode excited at $t=0$, $\beta(0)=\beta_0$, with no further reexcitations, $a_{\mathrm{in}}=0$. For $0<t<\tau$ nothing has yet returned from the mirror, $c_{\mathrm{in}}=0$, and Eq.~\eqref{eq:QED-b} gives exactly
\begin{equation}
\beta(t) = \beta_0\, e^{-(i\omega_c+\Gamma_0)t},\qquad 0<t<\tau ,
\label{eq:QED-free}
\end{equation}
with $\Gamma_0=(\gamma_{\mathrm{nr}}+\gamma_{\mathrm{rad}})/2$. This is the free superradiant decay of a 2DEG without a mirror. It does not depend on $d$, on $r_m$, or on the detuning of $\omega_c$ from the cavity modes. Iterating Eq.~\eqref{eq:QED-delay} gives the full solution as a series of echoes. In the frame rotating at $\omega_c$, $\beta(t)=\beta_0 B(t)e^{-i\omega_c t}$,
\begin{equation}
B(t) = \sum_{k\ge0} \Theta(t_k)\, \phi^k e^{-\Gamma_0 t_k} \sum_{j=0}^{k} C_{jk}\, (-\gamma_2)^j\, r^{\,k-j}\, \frac{t_k^{\,j}}{j!},
\label{eq:QED-series}
\end{equation}
with $t_k=t-k\tau$, $\phi=r_m e^{i\omega_c\tau}$, $C_{00}=1$, $C_{0k}=0$ for $k\ge1$ and $C_{jk}=\binom{k-1}{k-j}$ for $j\ge1$. In the second round trip, $\tau<t<2\tau$, $B(t)=e^{-\Gamma_0 t}-\gamma_2\phi\,(t-\tau)e^{-\Gamma_0(t-\tau)}$. On resonance, $\omega_c=(2m-1)\pi/\tau$ and $\phi=|r_m|$, so the matter amplitude has its first node at
\begin{equation}
t_1 = \tau + \frac{e^{-\Gamma_0\tau}}{\gamma_2 |r_m|},
\label{eq:QED-t1}
\end{equation}
which is $12.08$~ps (thick) and $5.27$~ps (thin), as in the main text. At long times $\beta(t)$ is dominated by the roots $s$ of the characteristic equation
\begin{equation}
\left(s+i\omega_c+\Gamma_0\right)\left[1-r\rho(s)\right]+\gamma_2\,\rho(s)=0,
\label{eq:QED-char}
\end{equation}
with $\rho(s)=r_m e^{-s\tau}$.
The two roots closest to the resonant Fabry-Perot mode are the lower and upper Landau polaritons (Tab.~\ref{tab:QED}).

\subsection{Occupations and the onset of reversible exchange}
\label{sec:QED-occ}
\begin{figure*}
    \centering
    \includegraphics[width=\linewidth]{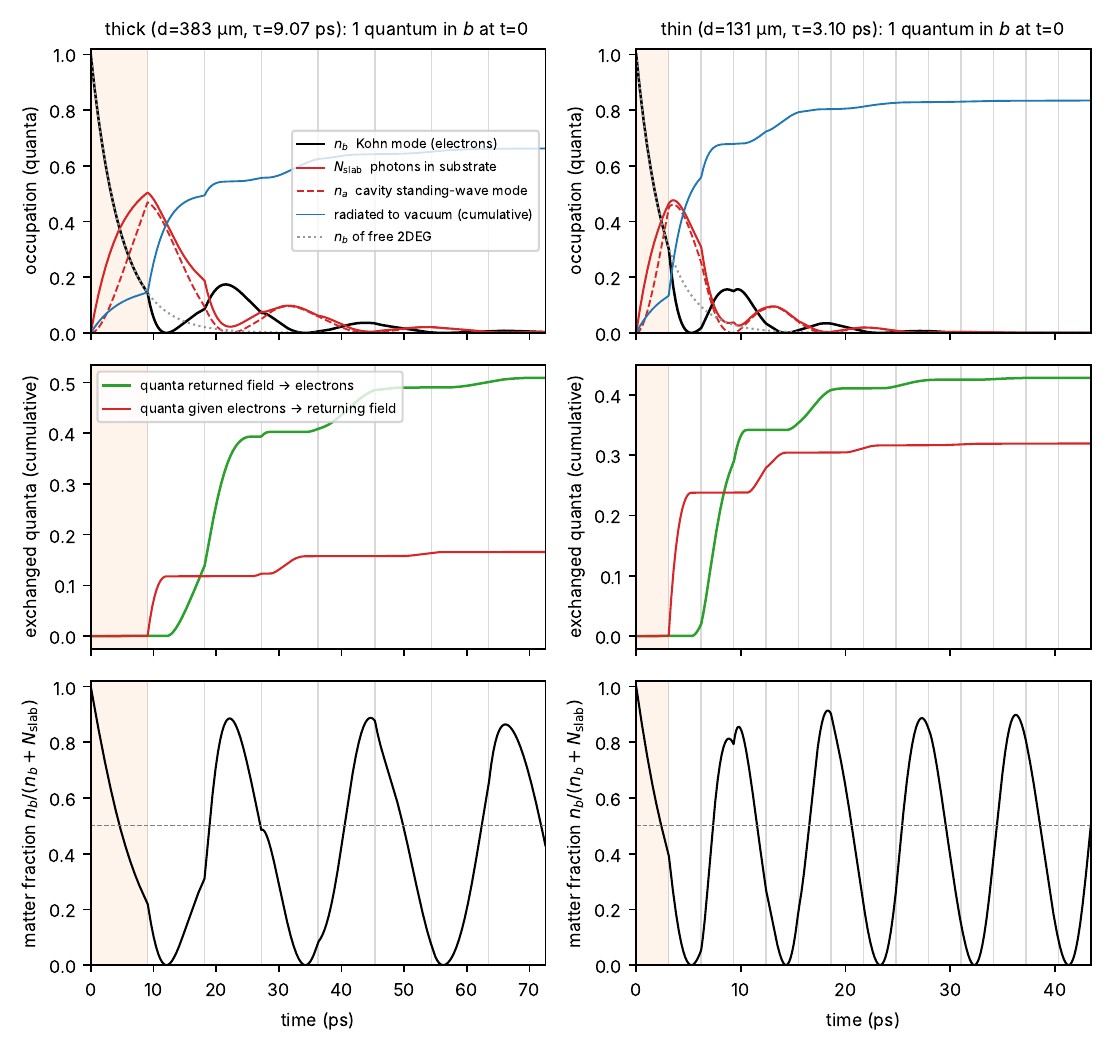}
    \caption {Quantum occupations of the thick (left column) and thin sample (right column) excited on resonance by a single-cycle THz pulse at $t = 0$, without. All quantities are normalized to the number of quanta the incident pulse delivers to the Kohn mode. The shaded region is the first cavity round trip, $0 < t < \tau$, and the grey vertical lines mark $t = k\tau$. Top: Kohn-mode (electron) occupation $n_b$, photons radiated by the electrons into the substrate $N_{\mathrm{slab}}$, their occupation $n_a$ of the resonant standing-wave cavity mode, quanta radiated into vacuum (cumulative), and $n_b$ of a free 2DEG without a mirror (dotted). Middle: cumulative number of quanta returned from the field to the electrons (green), given by the electrons to the returning field (red), and the net exchange (black), obtained from $P(t)$ of Eq.~(26). Bottom: matter fraction $n_b/(n_b + N_{\mathrm{slab}})$; the dashed line marks $1/2$.}
    \label{fig:occupations}
\end{figure*}
\begin{figure}
     \centering
     \includegraphics[width=\linewidth]{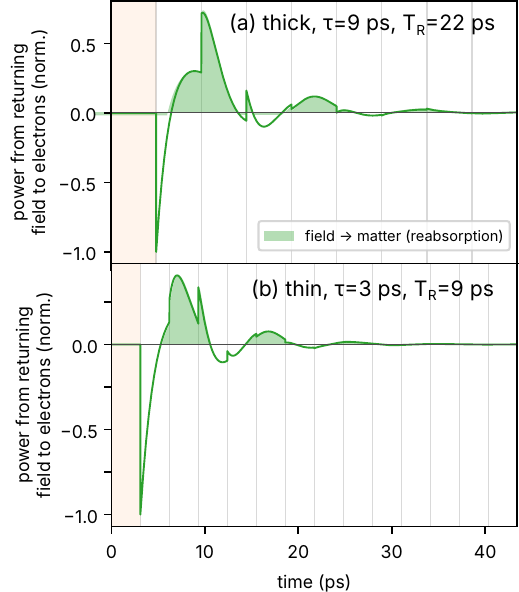}
\caption{Energy exchange $P(t)$ from the returning field $c_{out}$ to the CR Kohn's mode $b$.}
     \label{fig:QED-kick}
\end{figure}

We follow where a single quantum, put into the Kohn mode at $t=0$, goes. We define the electron (Kohn-mode) occupation $n_b(t)=|\beta(t)|^2/|\beta_0|^2$; the number of photons in the substrate $N_{\mathrm{slab}}(t)$, i.e.\ the flux $|c_{\mathrm{out}}|^2$ emitted into the substrate during the last round trip, with the mirror loss $1-|r_m|^2$ applied to the part already reflected; the occupation $n_a(t)$ of the standing-wave cavity mode $m$, obtained by projecting the substrate field onto the mode's profile; and the cumulative numbers of quanta radiated into vacuum ($|a_{\mathrm{out}}|^2$), lost to the electronic bath ($\gamma_{\mathrm{nr}} n_b$) and absorbed in the mirror. These add up to one at all times; numerically, the sum deviates from unity by less than $4\times10^{-4}$. From Eq.~\eqref{eq:QED-b},
\begin{equation}
\frac{dn_b}{dt} = -\gamma\, n_b + P(t),\qquad
P(t) = \frac{2\sqrt{\gamma_2}}{|\beta_0|^2}\,\mathrm{Re}\!\left[\beta^*(t)\, c_{\mathrm{in}}(t)\right].
\label{eq:QED-P}
\end{equation}
The first term is the irreversible loss of a free 2DEG. The exchange current $P(t)$ counts the quanta per unit time that the returning field gives back to the electrons ($P>0$) or extracts from them by stimulated emission ($P<0$). It is the term that drives the vacuum Rabi oscillations. Since $c_{\mathrm{in}}=0$ for $t<\tau$, $P$ vanishes identically in the first cavity round trip (Fig.~\ref{fig:QED-kick}).

\begin{table}
    \centering
    \begin{tabular}{lcc}\hline\hline
         & Thick & Thin\\\hline
        Kohn-mode occupation $n_b(\tau)$ & 0.144 & 0.307 \\
        $n_b(\tau)$ of a free 2DEG, $e^{-\gamma\tau}$ & 0.144 & 0.307 \\
        Photons in the substrate $N_{\mathrm{slab}}(\tau)$ & 0.50 & 0.46 \\
        Photons in the cavity mode $n_a(\tau)$ & 0.47 & 0.45 \\
        Radiated into vacuum by $\tau$ & 0.14 & 0.13 \\
        Matter fraction at $\tau$ & 0.22 & 0.40 \\
        Quanta returned to electrons, $0<t<\tau$ & 0 & 0 \\
        First increase of $n_b$ (ps) & 12.08 & 5.27 \\
        Quanta returned to electrons, $0<t<2\tau$ & 0.14 & 0.02 \\
        Quanta returned to electrons, all $t$ & 0.51 & 0.43 \\
        Mean matter fraction, $t>3\tau$ & 0.42 & 0.43 \\
        \hline\hline
    \end{tabular}
    \caption{Occupations per initial quantum in the Kohn mode, on resonance (Fig.~\ref{fig:occupations}).}
    \label{tab:occupations}
\end{table}

The results are shown in Fig.~\ref{fig:occupations} and Tab.~\ref{tab:occupations}. For $t<\tau$, the photon population in the slab increases, reaching about half a photon at $t=\tau$, almost entirely in the resonant cavity mode ($n_a=0.47$ for the thick sample). Simultaneously, the electron occupation follows the free-2DEG law $n_b=e^{-\gamma t}$ exactly; $dn_b/dt$ is never positive, and no quantum returns from the field to the electrons. The photons are in the cavity, but they are not hybridized with the electrons: the exchange runs one way only, from the electrons to the field, and is irreversible. After $\tau$, the returning field acts back on the electrons. Kohn mode $n_b$ first increases at $t_1$, Eq.~\eqref{eq:QED-t1}, and from then on $n_b$ and $N_{\mathrm{slab}}$ oscillate in antiphase. The matter fraction swings between about 0 and 0.9 around a mean of about 0.42, close to the Hopfield value $1/2$ of a resonant polariton~\cite{Hopfield1958}.

We therefore use the term vacuum Rabi oscillations for the reversible, periodic exchange of excitation between the Kohn mode and the cavity field, not merely for a beating of the measured signal. With this definition, the absence of vacuum Rabi oscillation in the first round trip is an exact statement of the model: the exchange current $P(t)$ vanishes identically for $t<\tau$. Polaritons, the normal modes of this reversible exchange, form only after one cavity round trip. This does not mean that the 2DEG is uncoupled from light before $\tau$; the coupling is what makes it radiate at the superradiant rate. What is missing before $\tau$ is the feedback from the cavity.

\newpage

\putbib[refs]


\end{bibunit}

\end{document}